\documentclass[aps,twocolumn,prb,superscript,floatfix,superscriptaddress,showpacs]{revtex4-1}
\usepackage{amssymb}

\usepackage{epsf}
\usepackage{epsfig}
\usepackage{graphicx}
\usepackage{dcolumn}
\usepackage{braket}
\usepackage{bm}
\usepackage{amsfonts}
\usepackage{amsmath}
\usepackage{amssymb}
\usepackage{color,soul}
\usepackage{wasysym}
\usepackage{mathrsfs}
\usepackage{natbib}
\usepackage{multirow}
\usepackage{times}

\newcommand{\bea}{\begin{eqnarray}}
\newcommand{\eea}{\end{eqnarray}}

\begin{document}

\title{Crafting zero-bias one-way transport of charge and spin}

\author{L. E. F. Foa Torres}
\altaffiliation{On leave from Instituto de F\'{\i}sica Enrique Gaviola (CONICET) and FaMAF, Universidad Nacional de C\'ordoba, Argentina.}
\affiliation{Departamento de F\'{\i}sica, Facultad de Ciencias F\'{\i}sicas y Matem\'aticas, Universidad de Chile, Santiago, Chile}
\author{V. Dal Lago}
\affiliation{Instituto de F\'{\i}sica Enrique Gaviola (CONICET) and FaMAF, Universidad Nacional de C\'ordoba, Argentina}
\author{E. Su\'arez Morell}
\affiliation{Departamento de F\'{\i}sica, Universidad T\'ecnica Federico Santa Mar\'{\i}a, Valpara\'{\i}so, Chile}

\begin{abstract}
We explore the electronic structure and transport properties of a metal on top of a (weakly coupled) two-dimensional topological insulator. Unlike the widely studied junctions between topological non-trivial materials, the systems studied here allow for a unique bandstructure and transport steering. First, states on the topological insulator layer may coexist with the gapless bulk and, second, the edge states on one edge can be selectively switched-off, thereby leading to nearly perfect directional transport of charge and spin even in the zero bias limit. We illustrate these phenomena for Bernal stacked bilayer graphene with Haldane or intrinsic spin-orbit terms and a perpendicular bias voltage. This opens a path for realizing directed transport in materials such as van der Waals heterostructures, monolayer and ultrathin topological insulators.
\end{abstract}
\pacs{ 73.20.At; 03.65.Vf; 72.80.Vp}

\date{\today}
\maketitle

\section{Introduction.}
More than ten years ago, the advent of graphene~\cite{Novoselov2004a,Novoselov2005a,Zhang2005,CastroNeto2009} kickstarted the discovery of the new family of two-dimensional (2D) materials. Today, they are used as building blocks for engineering a new way back to three-dimensions, the so-called van der Waals heterostructures.~\cite{Geim2013} A minimal unit with such hierarchy is a bilayer system composed of two weakly coupled layers, say $1$ and $2$. The general Hamiltonian has the form:
\begin{equation}
\label{H2}
{\cal H}= \left( \begin{array}{cc}
{\cal H}_1 & {\cal H}_{1,2} \\
{\cal H}_{2,1} & {\cal H}_2 \end{array} \right),
\end{equation}
where the intra-layer interactions encoded in the diagonal blocks dominate over the inter-layer coupling represented by the off-diagonal terms. Such arrangement may bring new possibilities which can already be noted in the case of bilayer graphene, the opening of a tunable bandgap by applying a perpendicular bias~\cite{Castro2007,Oostinga2008} being a noteworthy example. Moreover, the bandgap obtained in that way may even host marginal topological states.~\cite{Li2012}

Another type of bipartite system of interest includes two coupled topological insulators.~\cite{Hasan2010,Xiao2010,Ortmann2015} A typical junction involves two insulators with different topology~\cite{Shevtsov2012} side by side where robust edge states develop at the interface. A less studied case is that of a 2D topological insulator (TI) weakly coupled to a metal. Recently, it was shown that edge states may survive even on a gapless bulk.~\cite{Baum2015} Works addressing a non-equilibrium incarnation of TIs called Floquet topological insulators, have also pointed out that well defined topological states may exist, even on a gapless bulk.~\cite{Rudner2013,Perez-Piskunow2015,Titum2015}

Here we explore the properties of a 2D metal on top of a 2D topological insulator. Although the overall system has a gapless bulk, we find that edge states do coexist and those at one edge of the TI layer can be selectively \textit{switched-off} while the propagating states on the opposite edge do survive and may even keep their robustness to disorder and lattice imperfections. In a 2D TI such a selective switch-off is prevented by the bulk-boundary correspondence, a constraint which is circumvented in the composite metal on topological insulator (MOTI) system proposed here. As we will show below, the unique control of the edge states allowed by the MOTI sets the basis for one-way (directed) transport of charge and spin.

To illustrate this effect we consider bilayer graphene with either Haldane or intrinsic spin-orbit terms (ISO). Besides breaking inversion symmetry, a perpendicular bias voltage shifts the bands on each layer and can be used to generate an energy range where a 2D TI (say in the bottom layer) is weakly coupled to a metal (in the top layer). The selective switch-off of the edge states is evidenced as unpaired (non-reciprocal) edge states in the band structure which now bears a \textit{built-in} asymmetry between left and right moving states. By using a three-terminal setup where two leads are connected to the bottom layer and a third lead to the top one, the built-in asymmetry can then be exploited to generate nearly perfect one-way transport of charge or spin (depending on whereas a Haldane or ISO term is considered). Interestingly, transport occurs even \textit{without} a source drain bias, \textit{i.e.} it can be seen as a  pump effect operating without time-dependent potentials.

Two mechanisms are proposed for the selective switch-off of edge states. The first one exploits the strong sublattice polarization of the edge states together with the peculiar Bernal stacking. The stacking establishes a preferential coupling among those states polarized on the B sublattice of the TI layer and those in the A sublattice of the metal layer. In contrast, the set of states on the TI polarized on the A sublattice remain less affected by the top layer. The second mechanism is based on a geometrical setup where the top layer covers only one edge of the lower layer. While in the first case transport is fragile to perturbations, in the second case it improves when adding disorder or edge roughness, \textit{i.e.} it is \textit{antifragile}~\cite{Taleb2012} and thereby does not rely on the particular stacking order nor specifics of the model for the top layer, as long as it has a continuum able to hybridize with the topological states of the bottom layer. This motivates the acronym MOTI and opens the door for realizing similar phenomena in more general situations.

In the following we start by examining a spinless model of a graphene bilayer with a Haldane term. There we will see how the edge states can be selectively switched off and explain the underlying mechanism. Later on, in Sec. \ref{transport}, we discuss how this mechanism can be used to achieve perfect directional transport. Section \ref{spin} presents a results for a bilayer with intrinsic spin-orbit coupling, a case leading to spin and valley polarized currents. Finally, we discuss the possible experimental realizations of this proposal.

\begin{figure}[tbp]
\centering
\includegraphics[width=1.0\columnwidth, angle=0]{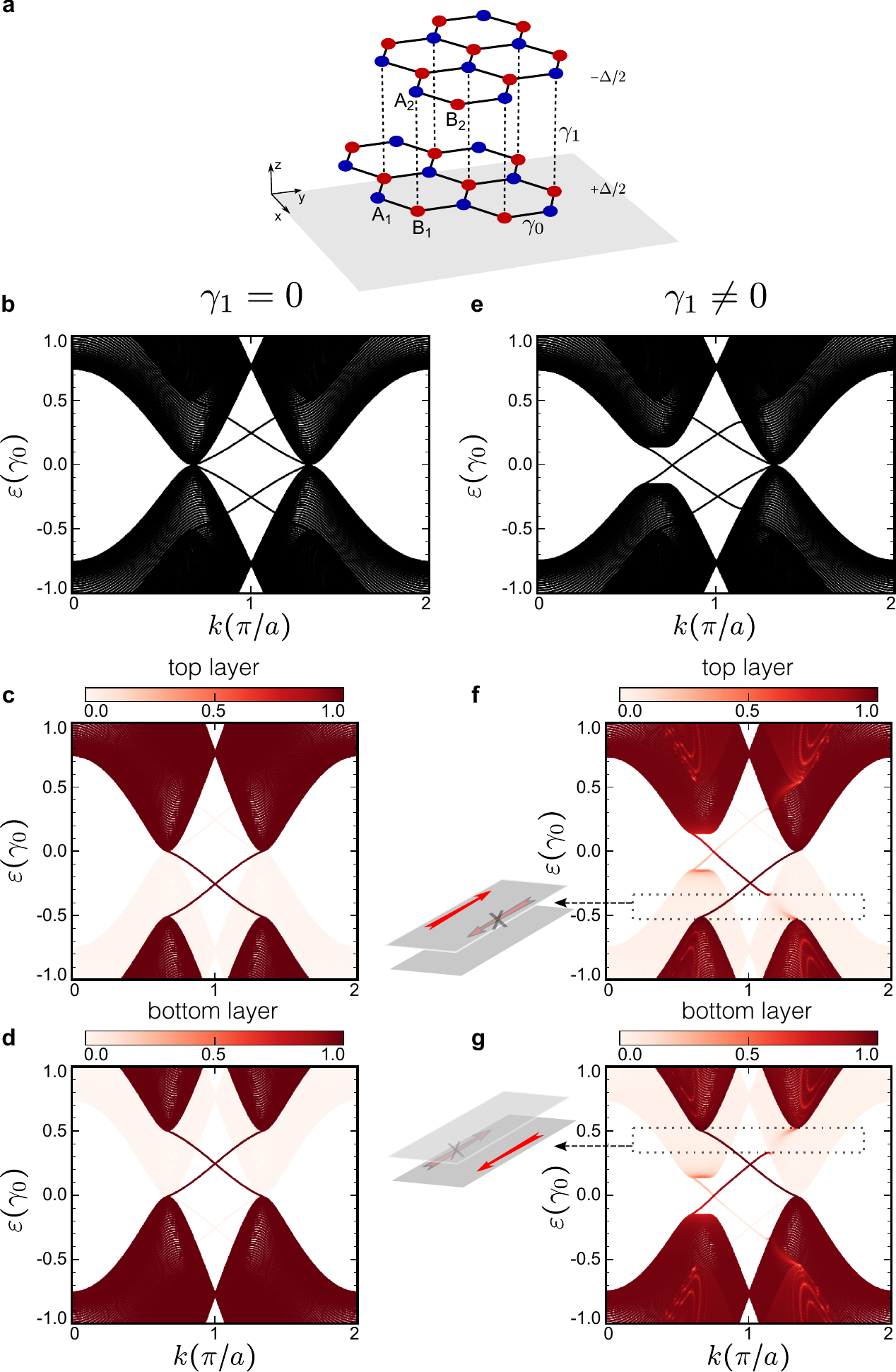}
\caption{(color online) (a) Scheme of a graphene bilayer with Bernal stacking and a bias voltage applied perpendicularly. (b) and (e) show the dispersion of the electronic states for a bilayer ribbon with a Haldane term and zig-zag edges. All the left panels correspond to $\gamma_1=0$ while the ones on the right are for $\gamma_1=0.15\gamma_0$. The same dispersion with a color scale encoding the weight of the corresponding states on each layer is shown in (c,d,f,g), thereby highlighting the regions where one of the edge states is selectively switched-off (dashed rectangles) once the interlayer interaction is turned on. The parameters in panels (b-h) are chosen with the aim of  illustrating the proposed mechanism, $\Delta=0.5\gamma_0$, $\gamma_{H}=-0.05\gamma_0$ and $W=104a$ ($a$ being the lattice constant).}
\label{fig1}
\end{figure}


\section{The biased Haldane bilayer: coexisting bulk and edge states.} 

To start with, let us analyze a graphene bilayer with a Haldane term.~\cite{Haldane1988} This spinless model will allow us to introduce the main ideas. Later on, we will consider a more realistic intrinsic spin-orbit (ISO) term. The Hamiltonian for spinless electrons is given by:
\begin{equation}
\label{Hamiltonian}
{\cal H}=\sum_{i}E_{i}^{{}}\,c_{i}^{\dagger}c_{i}^{{}}-\gamma_{0}\sum_{\left\langle i,j\right\rangle} c_{i} ^{\dagger}c_{j}^{{}}-{\rm i}\gamma_{H}\sum_{\left\langle\left\langle i,j\right\rangle\right\rangle }\nu_{i,j} c_{i} ^{\dagger}c_{j}^{{}}+{\cal H}_{\perp},
\end{equation}
where $c_{i}^{\dagger}$ and $c_{i}^{{}}$ are the electronic creation and annihilation operators at the $\pi$-orbital on site $i$,  $\gamma_{0}$ is the nearest-neighbors matrix element and $\gamma_{H}$ is the Haldane coupling. $\nu_{i,j}$ is $+1$ ($-1$) if the path from $j$ to $i$ is clockwise (anticlockwise). The on-site energies $E_{i}$ are chosen equal to $\Delta/2$ ($-\Delta/2$) for the sites on the lower layer (upper layer) as to model a perpendicular bias. ${\cal H}_{\perp}$ models the Bernal stacking (Fig. \ref{fig1}(a)) where $A_2$ sits on top of $B_1$ with the matrix element between them being $\gamma_{1}$.

\begin{figure}[tbp]
\centering
\includegraphics[width=0.85\columnwidth, angle=0]{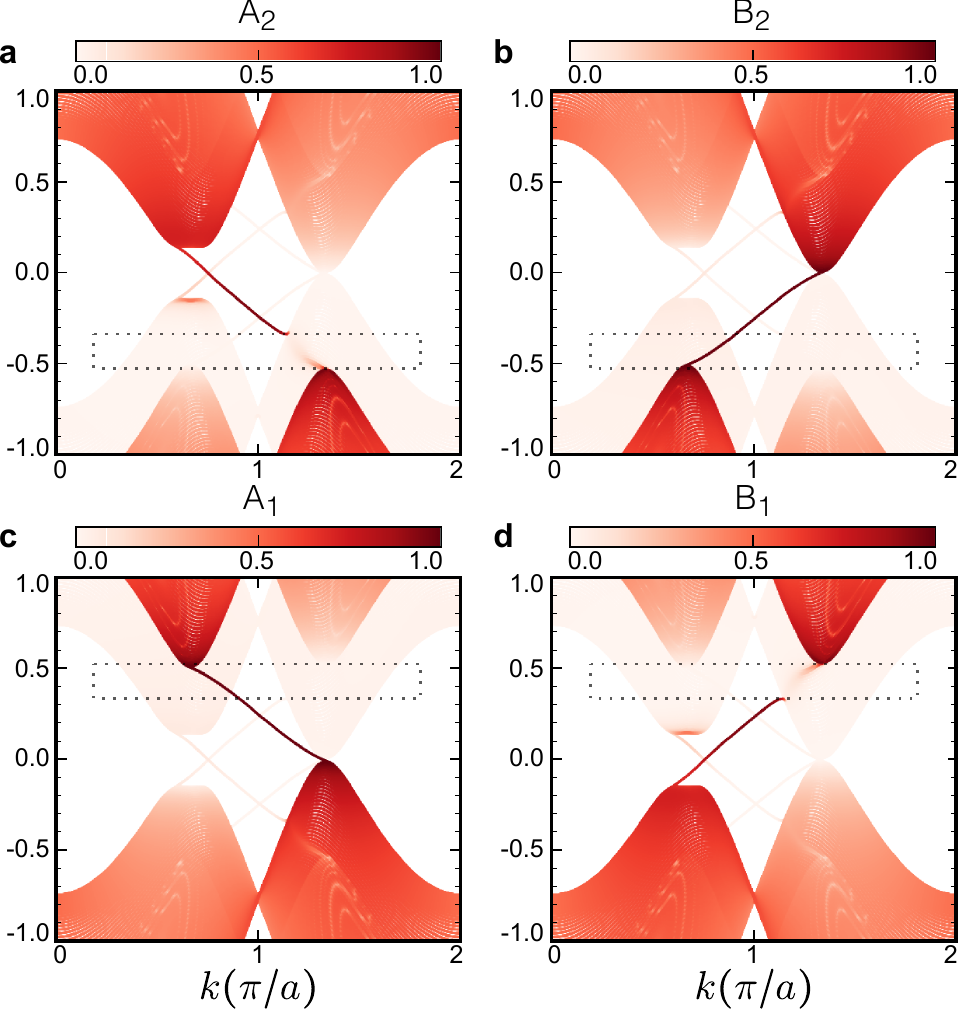}
\caption{(color online) (a--d) Dispersion relation resolved on each sublattice, the strong sublattice polarization of the edge states bridging the bulk gaps on each layer is evident. The parameters are the same as those in Fig. 1 panels (e--g).}
\label{fig2}
\end{figure}

The calculations presented here were carried out with home-made codes built on the Kwant~\cite{Groth2014} module. More details on the parameters used and additional results supporting our conclusions are available in the Appendices.

To motivate the discussion let us look at the bandstructure under a strong bias voltage $\Delta$ applied perpendicularly to the bilayer. A single of those layers bears a bulk gap with gapless Hall-like \textit{chiral} states propagating along the edges.~\cite{Haldane1988} If the interlayer interaction is turned off as in Fig. \ref{fig1} (b), the spectrum is the sum of that of each layer, that is, two copies of the same spectrum shifted by $\Delta$. When the interlayer interaction is turned on, it may be tempting to say that the main effect is to introduce a valley asymmetry close to zero energy as shown in Fig. \ref{fig1} (e), an effect which has been studied in detail before.~\cite{Gelderen2010} But, as we will see later on, there is a more striking asymmetry hidden below the gapless spectrum.

Figures \ref{fig1}(c--d) and (f--g) show the spectrum with a color scale encoding the weight of each state on the top (panel c and f) or bottom (panel d and g) layers, with (c-d) or without (f-g) interlayer coupling. Interestingly, in panels f and g there is a range of energies where one of the two chiral edge states is switched off (dashed rectangles).

The physical origin of this ''switch-off'' relies on the hybridization of the edge states with the continuum states on the other layer. Because of the large density of states on the ungapped layer, even a coupling $\gamma_1 \sim 0.1 \gamma_0$ is enough to produce new eigenstates which have a vanishing weight on the lower layer, thereby leading to the observed switch off. The missing link is now why the switch off happens to be selective and does not act equally on all the edge states.

To rationalize this, let us look at the spectrum aided by a color scale resolving the different sublattices (Figs. \ref{fig2}a--d). One can notice a strong sublattice polarization of the edge states with values of $99\%$. This is inherited from Haldane's model and assisted by the strong bias which deters mixing between layers. Moreover, the peculiar Bernal stacking (Fig. \ref{fig1}(a)) introduces a difference in the environment of those sites lying on top of each other ($A_2$ and $B_1$) and the others ($B_2$ and $A_1$): States polarized on $B_2$ or $A_1$ will be better protected from hybridization with the states in the other layer. \textit{Therefore, those edge states polarized on sublattices $A_2$ and $B_1$ are selectively switched-off by the coupling with the continuum on the other layer.} The valley where the switching off occurs depends on the bias voltage and the Haldane terms: \textit{switching the polarity of the bias or the sign of the $\gamma_{H}$ term, changes the valley on which the states are switched-off.}

Further results confirming the localization of the edge states of the composite system and the role of the interlayer bias are shown in Appendices \ref{AppA} and \ref{AppB} respectively.

\section{Harnessing the chiral edge state switch-off mechanism for achieving directional transport.} 
\label{transport}

\begin{figure}[t]
\includegraphics[width=0.90\columnwidth]{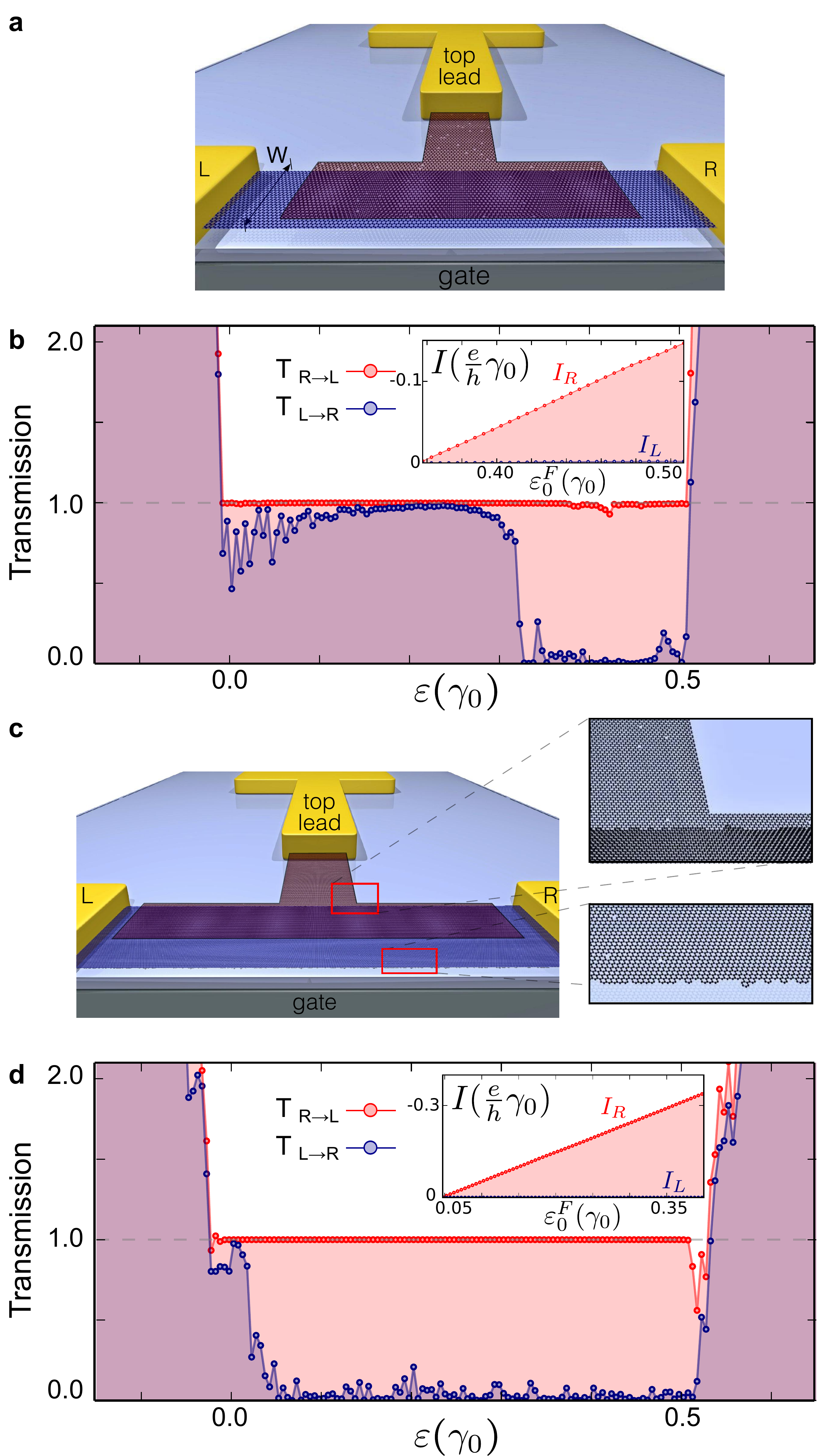}
\caption{(color online) (a) Scheme of a three-terminal geometry setup, where the top layer covers both edges. (b) Transmission probabilities between the left ($L$) and right ($R$) as a function of the electronic energy. Note the directional transport suppresion due to the switch-off of the edge state. The results in (d) are for the setup in (c), where the bottom layer is covered only partially (only one of the edges hybridizes with the continuum on top). Furthermore, to probe the robustness of this setup, roughness on the bottom layer and random disorder ($1\permil$ vacancies to the overall sample) are added. Parameters are chosen as in Fig. \ref{fig1}. To eliminate matching effects between the sample and the leads, the latter are chosen to be graphene monolayers with the same parameters ($\Delta$, etc.). The insets in (b) and (d) show the currents flowing on the left and right leads in the zero temperature limit. These currents are plotted as a function of the Fermi energies on $L$ and $R$ which are set to be equal to $\varepsilon_0^F$. The one on the top lead is: $\varepsilon_U^F=0.357\gamma_0$ (panel b) and $\varepsilon_U^F=0.0475\gamma_0$ (panel d).}
\label{fig3}
\end{figure}

Could we harness this mechanism to achieve directional transport? Connecting only two leads to one of the layers (left ($L$) and right ($R$)) does not provide the sought-for asymmetry $T_{L\rightarrow R}\neq T_{R\rightarrow L}$. Even though left-right symmetry is broken because time-reversal (TRS) and inversion (IS) symmetries are broken, the required probability current conservation applied to the corresponding $2\times2$ scattering matrix imposes $T_{L\rightarrow R} = T_{R\rightarrow L}$. To circumvent this problem, we use a trick inspired in the physics of quantum pumps~\cite{FoaTorres2005}: connecting an additional lead to the top layer (see Fig. \ref{fig3}(a)). Unitarity is now less restrictive on the resulting $3\times3$ scattering matrix, thereby permiting to obtain $T_{L\rightarrow R}\neq T_{R\rightarrow L}$ while allowing for independent control of the occupations.

As a proof of concept, we show in Fig. \ref{fig3}(b) the transmittances between the $L$ and $R$. Directional asymmetry sets in to allow for $T_{L\rightarrow R}\neq T_{R\rightarrow L}$ and, more interesting, when the edge states are selectively switched-off transport is almost perfectly \textit{non-reciprocal}.

So far we have relied on the sublattice polarization to achieve the selective switch-off. This has the disadvantage that, for example, changing the termination from zigzag to armchair will destroy the effect. Therefore, it would be desirable to improve it in such a way that: \textit{(i)} the switch-off is enhanced when adding disorder and edge roughness, and \textit{(ii)} the transporting edge states keep the robustness they had in the isolated monolayer. To do this, we propose a slightly different scheme based on uncovering one of the edges at the bottom layer as represented in Fig. \ref{fig3}(c). This effectively decouples the corresponding edge states from the continuum on the upper layer, thereby restoring their robustness against disorder and edge roughness. Moreover, the covered edge can still be effectively switched-off independently of the details of the termination, stacking order and specifics of the lattice on the top layer. This is illustrated in Fig. \ref{fig3}(d) where $T_{L\rightarrow R}$ and $T_{R\rightarrow L}$ are shown for a device $1\permil$ of vacancies distributed randomnly and rough edges. Interestingly, not only do the transporting states remain robust but also transport in the opposite direction is strongly suppressed in the \textit{full} span of the bulk gap.

One may wonder whether new edge states would be formed at the newly created monolayer-bilayer interface, but this is not the case as we have verified numerically. Two factors contribute to this behavior: each layer retains most of its properties because of the large bias, and the top layer is metallic preventing the application of the bulk boundary correspondence.

In the setup of Fig. \ref{fig3}(c) one observes a special feature: disorder and edge roughness tend to improve one-way transport as compared to a pristine system \footnote{An exception being a system with perfect zigzag termination as in Fig. \ref{fig1} where the edge state to be switched-off is polarized in the $A_1$ sublattice.} (further results are included in \ref{AppC}). This is different from merely being resilent to disorder as in the case of topological states and occurs because while the transporting edge states remain robust to disorder/roughness, the switching-off improves with it, \textit{i.e.} it is \textit{antifragile}.~\cite{Taleb2012}

We verified that the scattering matrix of the full system (having broken IS and TRS) is indeed non-reciprocal. But, is this enough to get a directed current at zero bias voltage? To answer this question one needs to consider the occupations at the different leads. The current flowing through lead $R$ is:

\begin{equation}
\label{current}
{\cal I}_R=\frac{e}{h} \int \sum_{\alpha \neq R}[T_{\alpha \rightarrow R}(\varepsilon) f_{\alpha}(\varepsilon)- T_{R \rightarrow \alpha}(\varepsilon) f_{R}(\varepsilon)] d\varepsilon,
\end{equation}
where $\alpha=L,U$, with $U$ being the top lead, and $f_{\alpha}$ is the Fermi-Dirac distribution at lead $\alpha$. A similar expression holds for the left lead. To get a directed currrent, besides the non-reciprocity of the scattering matrix, the chemical potential in the third lead ($\mu_{U}$) needs to be different from the one on the other layer ($\mu_0$), $\mu_{U}=\mu_{0}-\delta\mu$. Indeed, the transmittances obey the sum rule $\sum_{\alpha \neq R}T_{\alpha \rightarrow R}(\varepsilon)=\sum_{\alpha \neq R} T_{R \rightarrow \alpha}(\varepsilon)$ and therefore the kernel in Eq. (\ref{current}) is identically zero if all the occupations are equal. In the limit of perfect switch-off of one edge state, a current ${\cal I}_R=-(e/h)\delta\mu$ exits from the right lead towards the left one even at zero bias ($\mu_{L}-\mu_{R}=0$) (see Figs. \ref{fig3}(b) and (d) insets).

The incoming modes that would match with the switched-off states are not backscattered but rather diverted to the upper lead. Thus, the partial scattering matrix for the two lower leads is the one of an \textit{isolator}.~\cite{Jalas2013} When looking at the full scattering matrix we get a \textit{circulator},~\cite{Metelmann2015} charge flows from $R$ to $L$ to $U$ but not in the opposite direction. This is verified by explicit calculation of the currents (Figs. \ref{fig3}(b) and (d) insets) where we see that $I_L=0$ because the current injected by the right lead is compensated by a contribution of the same magnitude from $L$ to $U$.

\section{Biased bilayer with ISO: Pumping pure spin currents.}
\label{spin}
Let us now consider the Hamiltonian for a graphene bilayer with intrinsic spin-orbit (ISO) interaction~\cite{Gelderen2010,Prada2011}:
\begin{eqnarray}
\label{HamiltonianSO}
&&{\cal H}=\sum_{i, s_z}E_{i}^{{}}\,c_{i, s_z}^{\dagger}c_{i, s_z}^{{}}-\gamma_{0}\sum_{\left\langle i,j\right\rangle, s_z} c_{i, s_z} ^{\dagger}c_{j, s_z}^{{}}-\nonumber \\
&&-{\rm i}\gamma_{SO}\sum_{\left\langle\left\langle i,j\right\rangle\right\rangle, s_z }\nu_{i,j} s_z c_{i, s_z} ^{\dagger}c_{j, s_z}^{{}}+{\cal H}_{\perp},
\end{eqnarray}
where $c_{i, s_z}^{\dagger}$ and $c_{i, s_z}^{{}}$ are the electronic creation and annihilation operators at the $\pi$-orbital on site
$i$ with spin up $s_z=1$ or spin down $s_z=-1$.

\begin{figure}[tbp]
\includegraphics[width=0.95\columnwidth]{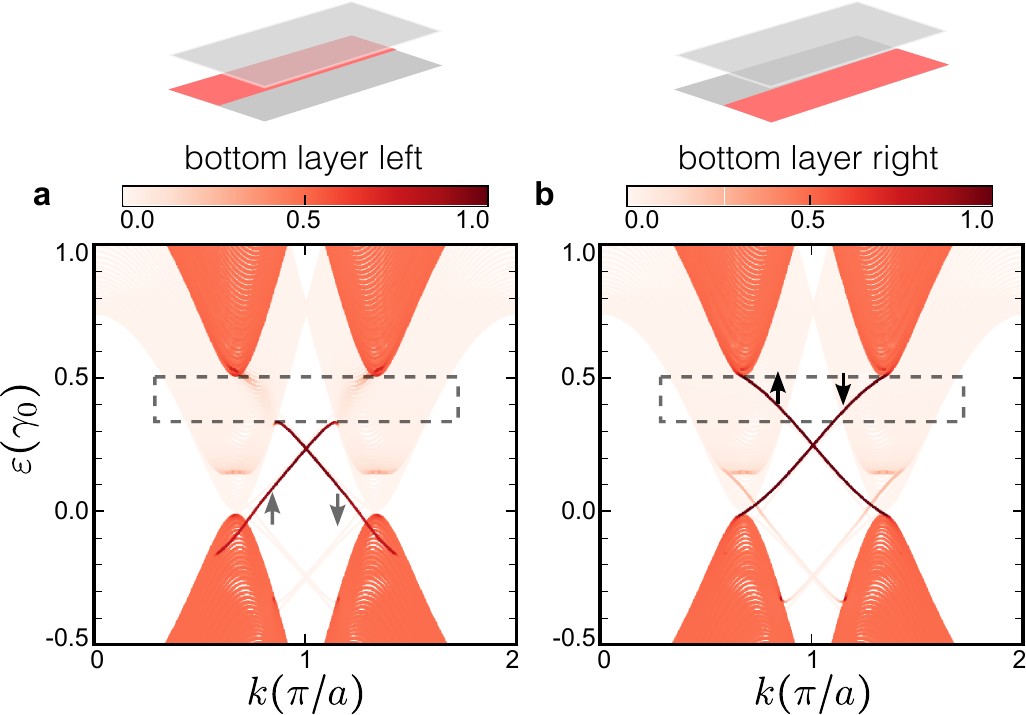}
\caption{(color online) (a) (b) Dispersion of the electronic states for a bilayer ribbon with an ISO term and zig-zag edges. The color scale corresponds to the weight of the corresponding states on the lower layer either on one half or the system as indicated by the schemes on top. Here, $\gamma_{SO}=-0.05\gamma_0$ while the other parameters are the same as in Fig. \ref{fig1} panels (e--g).}
\label{fig4}
\end{figure}

In the model of Eq. (\ref{HamiltonianSO}) we have two copies of the spinless case considered before, one for each spin $s_z$, but with an opposite phase in the second nearest neighbor term. Therefore, instead of having chiral edge states, one gets helical spin-polarized counter-propagating states.~\cite{Kane2005} Figure \ref{fig4} shows the spectrum for the spinful model with a color scale encoding the weight either on the left half or the right half of the lower layer (see shaded area in the schemes). The direction of the spin polarized edge states is also indicated with arrows. Although the required spin-orbit coupling is too small in graphene, the same physics can be realized in other systems such as silicene.~\cite{Ezawa2015,Tao2015}

Consistently with our previous discussion one sees that the selective edge state switch-off works this time destroying the edge states with both spin projections on the same border. Interestingly, we note that in a setup like the one shown in Fig. \ref{fig3}, one gets a pure spin and valley current (with vanishing charge current).

Such a setup would therefore pump spin and valley from the left lead into the right one and viceversa. The direction of the valley current is inverted by inverting the polarity of the perpendicular bias. In contrast to previous proposals for spin-injection where edge states in a non-topological phase where used,~\cite{Wimmer2008} here transport enjoys the robustness of the underlying topological phase.

\section{Summary and final remarks.}
Here we present a path for crafting a non-reciprocal band-structure with edge states propagating in only one direction in a finite system. To such end we exploit the effect of placing a metal (any system with a continuum spectrum in the energy range of interest) on top of a two-dimensional system which would otherwise be a topological insulator. We show that the hybridization between the edge states on the TI and the continuum on the metal, can then be harnessed to selectively switch off the states propagating on a chosen edge of the TI.

This can be done within the same material by using, for example, a perpendicular bias voltage as shown here for bilayer graphene with spin-orbit interaction. Though pristine bilayer graphene has a small spin-orbit coupling, the following physical realizations can be envisaged:

\textit{(i)} Since a stronger spin-orbit coupling is readily available in silicene or other two-dimensional materials~\cite{Ezawa2015,Tao2015} (which also have a honeycomb lattice), one can foresee a realization in a silicene sample with a graphene electrode on top;

\textit{(ii)}  Furthermore, even in graphene, there is theoretical and experimental evidence indicating that a spin-orbit coupling can be achieved through defects~\cite{Cresti2014} or by using a specific substrate as in Ref. ~\onlinecite{Calleja2015}. Therefore, an alternative would be using induced spin-orbit coupling on only one layer of a bilayer sample;

Another useful result is that by using a suitable geometry, the switch off mechanism has the feature of becoming more effective with added disorder and edge-roughness, \textit{i.e.} it is antifragile. An observable consequence of this effect is the generation of pure valley and spin directed currents, which can be detected through non-local transport measurements.~\cite{Valenzuela2006,Brune2012,Gorbachev2014} Our results may also find an application in ultrathin topological insulators,~\cite{Hong2010} and van der Waals heterostructures.~\cite{Geim2013}

\textit{Acknowlegdments.} LEFFT and VDL acknowledge financial support from SeCyT-UNC. LEFFT acknowledges the support of the Abdus Salam International Centre for Theoretical Physics (ICTP, Trieste). VDL thanks CONICET for the fellowship. ESM acknowledges Chilean FONDECYT grant Nr. 11130129. We thank Stephan Roche and Luis Brey for useful comments on an early version of this manuscript.

\textit{Author contributions.} ESM and LEFFT developed the concept and designed the study. LEFFT wrote a code and obtained the numerical results shown here, VDL verified the results. All authors discussed the results and agreed on the material to be selected for this work. LEFFT wrote the manuscript which was improved with comments and inputs from all authors.

\appendix

\section{Localization of the edge states}
\label{AppA}

\begin{figure}[h]
\centering
\includegraphics[width=0.95\columnwidth, angle=0]{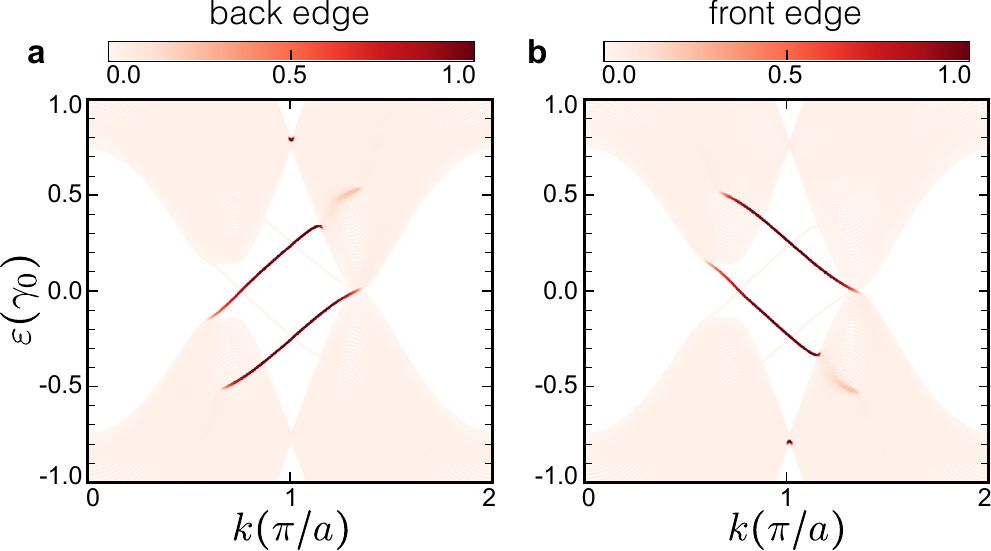}
\caption{(color online) Dispersion for the ribbon of Fig. 1 (e--g) of the main text, this time with a color scale encoding the weight of the states on a given edge (back (a) or front (b)).}
\label{figS1}
\end{figure}

The localization of the edge states in the model introduced in the main text can be appreciated in Fig. \ref{figS1}. The color scale encodes the weight of each state on sites up to $2.5a$ far away from the corresponding edge. The chirality of the edge states is evident.

\section{Role of the interlayer bias and other parameters used}
\label{AppB}

The results shown in the main text correspond to an interlayer bias of the same magnitude as the bulk gap of the unbiased system. Here we show complementary results for a situation where the interlayer bias is smaller $\Delta=0.3\gamma_0$. The selective edge state demolition takes place over a smaller energy range but it is otherwise not compromised.

The parameters used in the text were chosen to illustrate the basic idea. Even though bilayer graphene is known for having a small spin-orbit coupling, our model offers a simple and minimal situation to illustrate the undelying physical phenomenon which, we expect, could lead to further refinement and experiments in different materials and devices.

\begin{figure}[ptbh]
\vspace{0.5cm}
\centering
\includegraphics[width=0.95\columnwidth, angle=0]{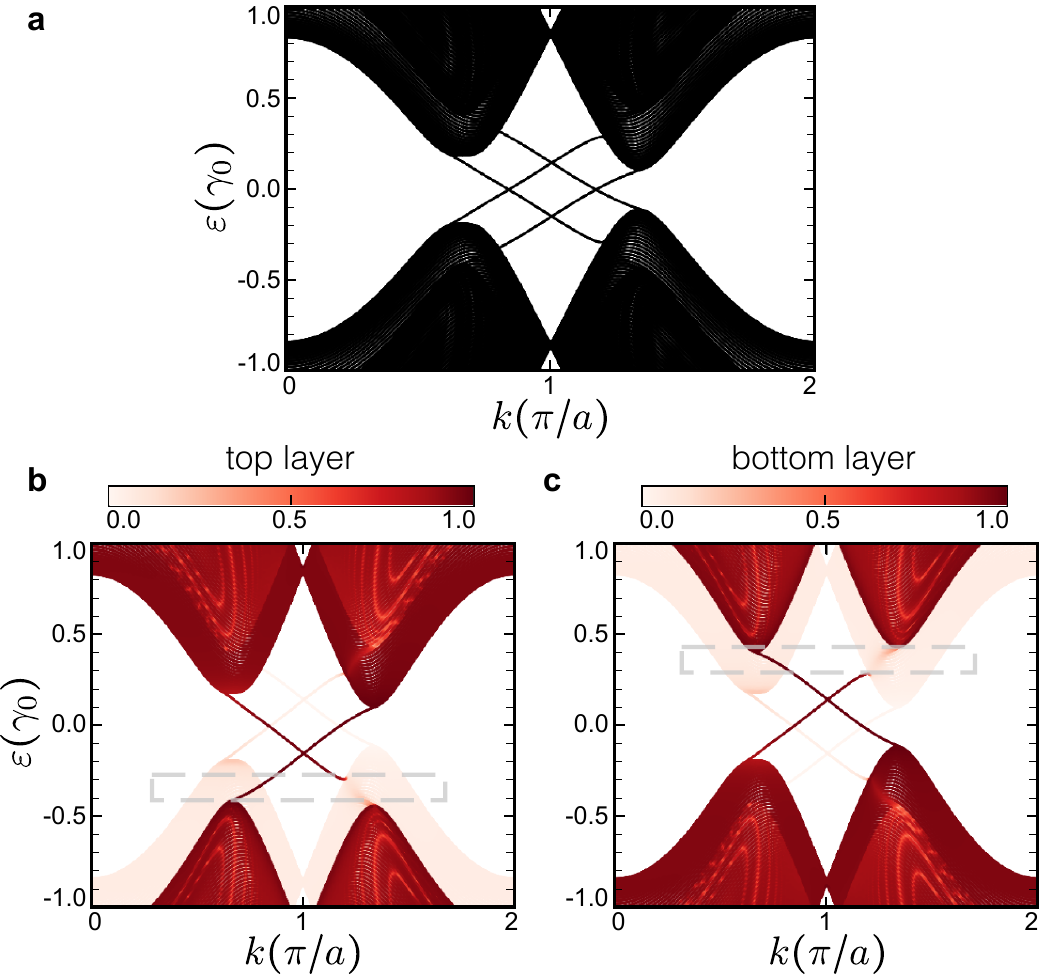}
\caption{(color online) (a) Dispersion for a ribbon under an interlayer bias of $\Delta=0.3\gamma_0$, the remaining parameters are the same as in Fig. 1 (e--g)of the main text. (b) and (c) show the weight of the corresponding states on the top and bottom layers respectively.}
\label{figS2}
\end{figure}

\section{Additional results for the transport properties}
\label{AppC}

\begin{figure}[tbph]
\centering
\includegraphics[width=0.95\columnwidth, angle=0]{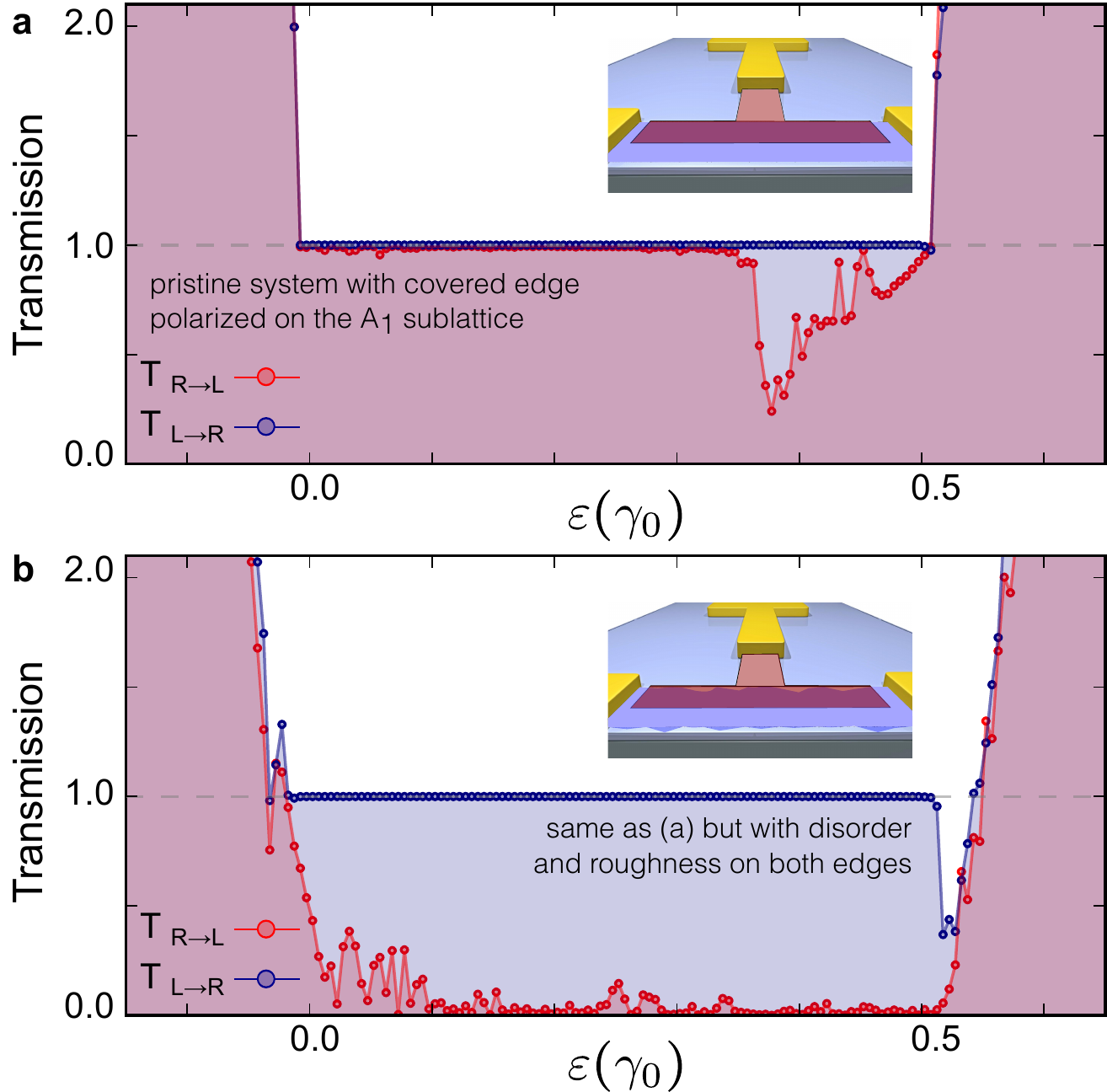}
\caption{(color online) Transmission probabilities (from $L$ to $R$ and viceversa) versus the energy of the incoming electrons. Both panels are for a case where only one edge of the lower layer is covered by the upper layer
and where the corresponding zigzag edge state is polarized on the $A_1$ sublattice. The demolition mechanism in this case is not fully effective for the pristine system (a) and is much improved when adding edge roughness and disorder (b). System parameters are chosen as in Fig. 3(d) of the main text.}
\label{figS3}
\end{figure}

In the discussion of Fig. 3 of the main text it was mentioned that edge roughness would generally improve the selective edge state demolition for the setup of Fig. 3(c). In that situation where one edge is covered, the demolition would indeed not be effective in the full energy if the system is pristine and the corresponding edge state is polarized on the $A_1$ sublattice. This is illustrated in Fig. \ref{figS3}(a). Adding edge roughness and disorder leads to an improvement of the one-way transport as evidenced in Fig. \ref{figS3}(b). Thus, the demolition mechanism is anti-fragile as it improves with disorder.

\noindent


%

\end{document}